\documentclass{ledger}
\usepackage[nice]{nicefrac}

\newcommand{\thefirstpagenum}[0]{1}
\newcommand{\thelastpagenum}[0]{5}
\newcommand{\theyear}[0]{201X}
\newcommand{\thevol}[0]{X}
\newcommand{\thedoi}[0]{DOI 10.5195/LEDGER.\theyear.X}
\newcommand{\ledgerpages}[0]{\thefirstpagenum-\thelastpagenum}

\hypersetup{pdfauthor={Bj\"orn Assmann, Samuel J. Burri}, pdftitle={Blockchain Scalability: A Linear Optimization Framework for Diversified Node Allocation in Shards}}

\title{Advancing Blockchain Scalability: \\
	A Linear Optimization Framework for Diversified Node Allocation in Shards}
\author{Bj\"orn Assmann\thanks{B. Assmann (bjoern.assmann@dfinity.org) is a senior researcher at the DFINITY Foundation, Zurich, Switzerland} Samuel J. Burri \thanks{S. Burri (samuel.burri@dfinity.org ) is vice president of engineering at the DFINITY Foundation, Zurich, Switzerland.}} 

\pretitle{
  \centering \selectfont 
  \centering \selectfont RESEARCH ARTICLE \par 
  \fontsize{24pt}{28pt}\selectfont} % Title is centered and at 24pt

\begin{document}
	
\maketitle

\thispagestyle{pagefirst}

\begin{abstract}
	Blockchain technology, while revolutionary in enabling decentralized transactions, faces scalability challenges as the ledger must be replicated across all nodes of the chain, limiting throughput and efficiency. Sharding, which divides the chain into smaller segments, called shards, offers a solution by enabling parallel transaction processing. However, sharding introduces new complexities, notably how to  allocate nodes to shards without compromising  the network’s security.

	This paper introduces a novel linear optimization framework for node allocation to shards that addresses decentralization constraints while minimizing resource consumption. In contrast to traditional methods that depend on random or trust-based assignments, our approach evaluates node characteristics, including ownership, hardware, and geographical distribution, and requires an explicit specification of decentralization targets with respect to these characteristics. By employing linear optimization, the framework identifies a resource-efficient node set meeting these targets. Adopted by the Internet Computer Protocol (ICP) community, this framework proves its utility in real-world blockchain applications. It provides a quantitative tool for node onboarding and offboarding decisions, balancing decentralization and resource considerations.	

%AUTHOR: keywords are OK to show for Review article, will be hidden for publication
%\begin{keywords}
%	\item Blockchain
%	\item Scalability
%	\item Database sharding
%	\item Linear optimization 
%	\item Node allocation 
%	\item Decentralization targets
%\end{keywords}
\end{abstract}

\section{Introduction}

		Blockchain technology has transformed the landscape of digital transactions and distributed systems. By replicating the ledger state among many nodes, the reliance on a single, central party is removed. This resulting decentralized trust model is attractive for a wide range of applications in fields such as decentralized finance, supply chain management, and cloud computing. While decentralization reduces trust assumptions, it introduces new scalability challenges. Replicating the ledger’s state across all participating blockchain nodes, can lead to bottlenecks affecting transaction throughput, latency, and increased energy consumption. 

To address scalability challenges in blockchain systems, sharding has become a well-established solution and is an active field of research \cite{Dang2019, Wang2019b,  Liu2022, Hashim2023}. Sharding, originally a database management technique, involves dividing a larger database into smaller, more manageable segments, or  `shards'. Each shard contains a subset of the total data, enabling parallel processing and thus significantly improving response times and throughput. This concept is well-illustrated in the context of traditional database systems, such as in Facebook's sharding implementation \cite{Annamalai2018}. In a blockchain context, sharding extends this concept to distribute the network's transactional load across multiple, smaller sets of nodes. Each shard processes a portion of the total transactions, thereby reducing the burden on individual nodes and enhancing the overall capacity of the system. This parallel processing capability is crucial for blockchain scalability, as it directly addresses the bottlenecks associated with transaction throughput and latency in a growing network.

However, implementing sharding in blockchain contexts introduces unique challenges, particularly around security. One of the primary security vulnerabilities in a sharded blockchain is the risk of a single-shard takeover attack. In such an attack, malicious actors may gain control of a single shard potentially compromising the integrity of transactions or data within that shard. Because a shard is only composed of a subset of the network’s nodes, fewer actors may be required to compromise a shard than for compromising the entire network. In other words, shards may weaken a network’s decentralization and thus the allocation of nodes to shards is essential for a secure sharding solution. The allocation strategy must consider the diversity and trustworthiness of nodes within each shard to prevent collusion and ensure robustness.

This paper contributes to the evolving discussion on blockchain sharding by introducing a new modeling framework focused on the allocation of nodes to shards. Traditional methods in the literature for node allocation to shards often rely on random assignments of nodes to shards \cite{Liu2022} in conjunction with  trust-based  \cite{Yun2019, Zhang2023} or performance-based  \cite{Wang2019} distribution schemes. In contrast to that, we introduce several novel concepts for the allocation of nodes to shards, thereby enhancing the granularity and effectiveness of decentralization assessments in blockchain networks.

First, we introduce the concept of \emph{node characteristics} to provide a more nuanced evaluation of decentralization. This approach allows to specify decentralization requirements for multiple aspects such as node ownership, data center control, geographic location, jurisdiction, and hardware configurations. This multifaceted perspective expands upon traditional blockchain decentralization assessments, which often focus solely on node ownership. For instance, by considering the characteristic 'country', distributing nodes across multiple countries within a shard mitigates the risk of disruptions due to environmental, regulatory, or legal changes specific to a single country.

Second, we establish the concept of a \emph{target topology} for the network that specifies  the number of shards, their sizes, alongside decentralization targets for each shard. This approach allows for tailored decentralization strategies, aligning the level of decentralization with the intended use and criticality of each shard.

Third, we employ a linear optimization model to determine an allocation of nodes to shards that meets the target topology with minimal resource usage. This optimization balances security needs with resource constraints. Through iterative evaluations of various target topology options and their associated resource requirements, blockchain networks can make well-informed decisions on the most suitable target topology to adopt.

This model has been implemented in Python \cite{topology_optimizer_2023} and validated through its application in the Internet Computer Protocol (ICP), a live blockchain platform \cite{Dfinity2022}.  We provide real-world examples of a target topology and visualize the optimization output, demonstrating the model's efficacy in enhancing blockchain scalability and viability.

\newpage

The rest of this paper is organized as follows. Section 2, ``Metrics of Decentralization of Nodes in a Shard,'' introduces the Nakamoto Coefficient and the concept of a shard limit to measure node decentralization. Section 3, ``Modeling Approach,'' outlines our linear optimization model for determining an optimal balance between the number of required nodes and network decentralization. In Section 4, ``Modeling Results,'' we present the aforementioned case study on the Internet Computer Protocol (ICP), applying our model to real-world data to illustrate its effectiveness in optimizing node allocations. Finally, the paper concludes in Section 5 with a discussion of our main findings and future directions for research.

	%%%%%%%%%%%%%%%%%%%%%%%%%%%%%%%%%%%%%%%%%%%%%%%%%%%
\section{Metrics of Decentralization of Nodes in a Shard}

The Nakamoto Coefficient \cite{Lin2021} traditionally signifies the smallest number of entities required to control more than 50\% of resources in a decentralized system. It provides a measure of how decentralized the control of a system is; the higher the Nakamoto Coefficient, the more decentralized a system is.

The threshold, above which malicious nodes can stall or corrupt a shard, depends on the blockchain protocol’s consensus algorithm. Discussions of different consensus algorithms are outside the scope of this paper. We thus use a generic threshold variable $t$ so that our approach is suitable for all consensus algorithms, defining the Nakamoto Coefficient as the smallest subset of entities corresponding to a specific type of characteristic (such as node owner, data center, country, etc.) that collectively control at least the threshold $t$ of the nodes in a shard. 

More formally, we define the \emph{Nakamoto Coefficient for a shard} as follows. Given
\begin{itemize}
	\item {$n$} is the \emph{number of nodes} in the shard,
	\item {$\chi$} is a \emph{characteristic} of the nodes assuming one of the values in $\{ \chi_1, \chi_2, … \}$,  
	\item {$\xi(\chi_i)$} denotes the number of nodes of the shard with the characteristic value {$\chi_i$}, and
	\item {$t$} is the threshold of nodes required to stall or corrupt a shard.
\end{itemize}
Then the Nakamoto Coefficient for the characteristic $\chi$ is defined as:
\[ NC_{\chi}(n) = \min |P| \]
where $P \subseteq \chi $  such that:
\[ \sum_{\chi_i \in P} \xi(\chi_i) \geq t n \]

Let’s look at an example with a shard composed of $n = 7$ nodes and let’s assume the blockchain protocol requires a threshold of $t=\nicefrac{1}{3}$ nodes to stall a shard. We want to determine the Nakamoto Coefficient with respect to the characteristic country $ \chi = \{ us, uk, ch, de \}$ in which these nodes are operated, i.e. how many countries need to collude to stall the shard? Assuming $\xi(us) = 2$, $\xi(uk) = 1$, $\xi(ch) = 2$, and $\xi(de) = 2$, it follows that $NC_{\chi}(n) = 2$. There is no single country $\chi_i$ that controls the shard, no $\{ \chi_i \} \subset \chi$ such that $\xi(\chi_i) \geq \nicefrac{7}{3}$. However, any two countries can control the shard. For example for $P = \{us, uk\}$, $\sum_{\chi_i \in P} \xi(\chi_i) = \xi(us) + \xi(uk) = 2 + 1 = 3 \geq \nicefrac{7}{3}$.

A simpler, yet less nuanced, node decentralization constraint is the \emph{shard limit}. It imposes a constraint on the maximum number of nodes within a shard that can share the same characteristic value $\chi_i$. Formally, a shard respects a shard limit $l$ with respect to a characteristic $\chi$ if, for any  $ \chi_i  \in \chi $, there are no more than $ l$ nodes in the shard with the value  $\chi_i$. Using the notation above, this constraint is satisfied when $ \xi(\chi_i) \leq l $ for every $ \chi_i \in \chi $. 
 As an illustrative example, if the shard limit for the characteristic 'node owner' is set to 1, then each node owner must be unique within the shard; no two nodes in the shard are owned by the same entity.

The relationship between the Nakamoto Coefficient and shard limit is direct: If each node in a shard has a different value of a characteristic $\chi$, i.e., the shard respects the shard limit = 1, then the Nakamoto Coefficient $NC_{\chi}(n)$ is $\lceil t  n \rceil$ \footnote{$\lceil x \rceil$ is the ceiling function that maps $x$ to the smallest integer $i$ such that $x \leq i$}, which  is the maximum Nakamoto Coefficient achievable. For example, assuming that the number of nodes in a shard is $n = 13$ and that the threshold $t=\nicefrac{1}{3}$, the maximal Nakamoto Coefficient is $\lceil \nicefrac{13}{3} \rceil = 5$.   

If each characteristic value may appear up to twice in a given shard, i.e., the shard respects the shard limit = 2, the Nakamoto Coefficient will be reduced. In this case, the Nakamoto Coefficient $NC_{\chi}(n)$ is $\lceil \frac{tn}{2} \rceil$. The factor $\nicefrac{1}{2}$ accounts for the fact that now, to reach the same threshold of control (e.g., $\nicefrac{1}{3}$ of the nodes), fewer unique characteristic values are needed, as each characteristic value  can appear twice. Taking the 13-node shard as an example again, the Nakamoto Coefficient is now $\lceil \nicefrac{13}{6} \rceil = 3$.  In general, for a shard limit \(L\), the Nakamoto Coefficient \(NC_{\chi}(n)\) can be determined using the formula \(NC_{\chi}(n) = \lceil \frac{tn}{L} \rceil\).

\section{Modeling Approach}
In this section, we present a mathematical model for determining an optimal balance between the number of required nodes and network decentralization. We employ linear optimization \cite{Bertsimas1997}, describing the model input, applied constraints, and objective functions. 

Results of the model will be discussed in the subsequent section. An implementation of the model in  Python is publicly available on Github\cite{topology_optimizer_2023}. 

\subsection{Input}
\emph{Nodes and Their Characteristics:} This represents the  nodes and their characteristics. Every node is either existing, i.e. already used by the network, or it can be added to the network, which would increase resource utilization. \\
\emph{Target Topology:} This describes the desired structure of the network, detailing the count and size of shards. For each characteristic, a decentralization goal is specified, which could be either the target Nakamoto coefficient or a shard limit. 

Table~\ref{example_shard_structure} shows an example target topology including decentralization constraints for the node characteristic country. In this example topology,  shards are categorized into two types based on their significance to the network's security and integrity: `Critical' and `regular'. Critical shards are those whose operation is essential for maintaining the network's core functions and overall security, whereas regular shards handle less sensitive operations. 

\vspace{10pt}
\begin{table}[h]
	\centering
	\begin{tabular}{|l|c|c|c|c|}
		\hline
		\textbf{Shard type} & \textbf{\# Shards} & \textbf{\# Nodes} & \textbf{Country limit} \\
		\hline
		Critical  & 2 & 43 &  2 \\
		\hline
		Regular  & 10 & 13 &  3 \\
		\hline
	\end{tabular}
	\caption{Example target topology.}
	\label{example_shard_structure}
\end{table}
\vspace{10pt}

\subsection{Constraints}
Two types of constraints are considered: \emph{Nodes to Shard Allocation Constraints} and \emph{Characteristic Constraints}. These constraints are defined using matrices that capture the relationship between nodes, shards, and node characteristics.
\newpage 
\emph{Nodes to Shard Allocation Constraints}
Let $ A_{ij} $ be a binary matrix where the index $i$ represents nodes and the index $j$ represents shards. This matrix maps nodes to shards, where $A_{ij} = 1$ indicates that node $i$ has been allocated to shard $j$, while $A_{ij} = 0 $ indicates no allocation. The matrix adheres to the following constraints:

\begin{itemize}
	\item {\emph{Uniqueness}}: Each node can be allocated to exactly one shard or not allocated at all. This means, for each node $ i $, $ \sum_{j} A_{ij} \leq 1 $.
	
	\item {\emph{Shard Fill}}: The size of each shard $S(j)$ should match the number of nodes assigned to it. This means, for each shard $ j $,  $ \sum_{i} A_{ij} = S(j) $.
\end{itemize}

\emph{Characteristic Constraints:}
For each node characteristic $\chi$, e.g., node owner, a characteristic matrix $ C_{kj} $ is constructed, where $ k $ indexes the characteristic values, and $ j $ indexes the shards. This matrix is derived from the node-to-shard allocation matrix $ A_{ij} $. If $\chi(i)$ denotes the characteristic value of a node $i$, then $ C_{kj}  = \sum_{i} A_{ij} \cdot \mathbf{1}_{\{\chi(i) = k\}} $, where $\mathbf{1}_{\{\chi(i) = k\}}$ is the indicator function that is 1 if $\chi(i) = k$ and 0 otherwise. This matrix must satisfy one of the following constraints:

\begin{itemize}
	\item {\emph{Shard Limit}}: If a shard limit $ l $ is set for a characteristic, then for each characteristic value $ k $ and shard $ j $, $ C_{kj} \leq l $. For instance, if the shard limit for the characteristic node owner is~2, then no more than two nodes with the same owner can be allocated to any single shard.
	
	\item {\emph{Nakamoto Coefficient}}: Given a Nakamoto coefficient target NC, ensure that for no subset of characteristic values smaller than NC, the sum of the number of nodes with these values are above or equal the threshold $t \cdot n(j)$, where $n(j)$ denotes the number of nodes in shard $j$. Formally, for all subsets $P \subset \chi$ with $|P| < \text{NC}$, it must hold that $\sum_{k \in P} C_{kj} < t \cdot n(j)$. For instance, if $\text{NC} = 5$ and $t=\nicefrac{1}{3}$, no subset of nodes owned by 4 owners controls a third of the nodes in the shard.
\end{itemize}

Please note that using the Nakamoto Coefficient leads to a greater number of constraints and hence, from a computational performance perspective, it is preferable to use the shard limit when possible.

\subsection{Objective Function}
The linear optimization model presented in this paper has two distinct optimization strategies, each with its own objective and application context. 

%\subsubsection{Minimizing Additional Nodes}
\emph{Minimizing Additional Nodes}: The  objective under this strategy is to minimize the number of additional nodes required to achieve the specified target topology. This approach is particularly relevant when expanding the network, where the goal is to find the most resource-efficient way to enhance network decentralization without incurring unnecessary costs. 

%\subsubsection{Maximizing Decentralization with Existing Nodes}
\emph{Maximizing Decentralization with Existing Nodes}: In scenarios where the addition of new nodes is not planned or feasible, an alternative optimization strategy can be pursued. 
This strategy focuses on the optimal allocation of the existing nodes across the network's shards,  to achieve the highest possible level of decentralization. 
Decentralization here is quantitatively assessed by the aggregate of Nakamoto coefficients across the shards. Additionally, the model allows for the application of weightings, providing flexibility to prioritize specific characteristics or shards based on strategic objectives or network requirements.

\subsection{Model Application}
This section elaborates on the practical applications of the model in evaluating node allocation strategies and assessing the potential impact of node candidates on the network's decentralization. 
\newpage
%	\subsubsection{Assessment of Status Quo}
\emph{Assessment of Status Quo}: Utilizing the objective function for maximizing decentralization with existing nodes, the model assesses the current level of decentralization achievable with the existing node set. This evaluation provides a baseline understanding of the network's decentralization status without the introduction of new nodes. 

%	\subsubsection{Iterative Assessment of Target Topologies}
\emph{Iterative Assessment of Target Topologies}: Through the objective function of minimizing additional nodes, the model facilitates an iterative assessment of various target topology candidates. This process involves estimating the required number and type of additional nodes to meet each proposed topology. By simulating multiple topology scenarios, the model aids in determining the most resource-efficient and effective configuration for achieving desired decentralization levels. 

%	\subsubsection{Evaluating Node Candidates}
\emph{Evaluating Node Candidates}: 	Given a set of existing nodes and a set of candidate nodes that may be added, the model can be used to assess:
\begin{itemize}

	\item Effectiveness of Candidate Nodes: Determine whether a candidate node can contribute towards the target topology. Specifically, a node is considered to contribute to the target topology if its inclusion decreases the number of further nodes required by one.
	Note that this methodology is designed to prevent the occurrence of a local minima as the following observation illustrates:
	Let $N$ be the set of current nodes. Furthermore, let $k$ be the minimal number of additional nodes required to reach the target topology as computed by the model.  Should adding an additional node $n$ to the node set $N$ decrease the count of needed additional nodes by exactly one, it implies that for the augmented set  
	$N \cup \{n\}$, the number of additional nodes necessary drops to $k-1$.
	Furthermore, as by the design of the model, there exist at least one set of $k-1$ additional nodes that, once added to $N \cup \{n\}$, would satisfy the target topology.
	\item Relevance of Existing Nodes: Identify which existing nodes are not required for achieving the target topology and could potentially be removed.
\end{itemize}

\section{Modeling Results}

In this section, we illustrate the application of the presented model to a real-world use case, using the nodes and shards of the Internet Computer Protocol (ICP) as an example. The ICP consensus algorithm  \cite{Dfinity2022} guarantees  safety and liveness, in the presence of less than one-third (i.e., $t=\nicefrac{1}{3}$) of faulty or malicious (Byzantine) nodes.
We apply the objective function ‘Minimizing Additional Nodes’ as described in the previous section.

\subsection{Utilized Model Inputs}
\label{sec:utilized_model_inputs}
Table~\ref{target_shard_structure}  specifies the number, types, and sizes of anticipated shards agreed upon by the ICP DAO\cite{ICProposalNov2023}, called Network Nervous System (NNS). The additional column labeled `Gen2' indicates whether the shard is designated to run on generation two machines, enhancing protection against malicious actors. This is an additional constraint, which is considered in the optimization scheme. The number of shards is based on current and anticipated demand. The sizes of the shards were chosen depending on the sensitivity of the services or dapps running on them. For instance, the NNS shard, hosting the ICP DAO, has the highest sensitivity and is thus proposed to have the largest number of nodes allocated. 
%\vspace{10pt}

\begingroup
\setlength{\intextsep}{100pt} % Adjusts the space above and below the table
\begin{table}[h]
	 \captionsetup{belowskip=10pt}
	\centering
	\begin{tabular}{|l|c|c|c|c|}
		\hline
		\textbf{Shard type} & \textbf{\# Shards} & \textbf{\# Nodes} & \textbf{Total} & \textbf{Gen2} \\
		\hline
		NNS & 1 & 43 & 43 & no \\
		\hline
		SNS & 1 & 34 & 34 & no \\
		\hline
		Fiduciary & 1 & 28 & 28 & no \\
		\hline
		Internet Identity & 1 & 28 & 28 & yes \\
		\hline
		ECDSA signing & 1 & 28 & 28 & yes \\
		\hline
		ECDSA backup & 1 & 28 & 28 & yes \\
		\hline
		Bitcoin canister & 1 & 13 & 13 & no \\
		\hline
		Application shard (Gen2) & 2 & 13 & 13 & yes \\
		\hline
		Application shard & 31 & 13 & 403 & no \\
		\hline
		Reserve nodes & & & 120 & \\
		\hline \hline
		\textbf{Total} & & & \textbf{751} & \\
		\hline
	\end{tabular}
	\caption{Suggested target shard structure.}
	\label{target_shard_structure}
\end{table}
\endgroup
%\vspace{10pt}
\newpage
The currently available set of nodes is extracted from the ICP dashboard as of September 7, 2023, totals 1151 nodes, with 84 of them being generation two nodes.
The analysis incorporates candidate nodes from new countries, data centers, and data center providers. For every new country, we assume that there are up to  20 additional nodes (5 node owners x 4 nodes per node owner), and these candidate nodes are generation 2 machines.

The process of collecting data on node characteristics is initiated during the node onboarding process, where node owners, who are not anonymous, submit extensive documentation. This documentation includes a self-declaration comprising a statement of identity, a statement of provision of node machines, and a statement of good intent, alongside a proof-of-identity document.

Following the submission of these documents, the node owner presents their case, including the node owner name, data center, data center provider and country, on a designated forum, stimulating community discussion and scrutiny. Subsequently, node owners submit a formal proposal to the Internet Computer DAO, including hashes of the previously submitted documentation, reinforcing data integrity and traceability. The DAO participants then vote on the node owner onboarding proposal and separately on the node onboarding proposals. If approved, the corresponding node characteristics are stored on the blockchain. 

There are techniques to automatically verify the correctness of characteristics and thereby reduce trust assumptions. For example, to approximate the geographic location of these nodes after onboarding, two methods can be employed: analyzing the IPv6 addresses and measuring round-trip times. While IPv6 addresses can provide initial geographic indications, they are not infallible due to the potential use of VPNs to alter perceived locations. Therefore, measuring round-trip times serves as an additional verification method, offering a more nuanced approach to determining node locations.

\subsection{Model Results Optimizing Node Allocation}
\label{sec:model_results}
In this section, we detail the outcomes of performing a sensitivity analysis through the application of our optimization model to a sequence of decentralization targets,
using  shard limits to specify the decentralization constraints. Our model simultaneously considers all node characteristics, optimizing for a holistic and integrated approach to decentralization across multiple dimensions.
Based on the initial assessment of node decentralization discussed in the appendix, we choose the following series of decentralization targets, each with increasing levels of stringency.

\begin{itemize}
	\item {\emph{Shard Limit 3}}: This target sets a shard limit of 3 for all node characteristics, meaning no more than three nodes with the same characteristic value are allocated within a shard.
	
	\item {\emph{Shard Limit 2}}: This target imposes a more stringent constraint, limiting each characteristic value to a maximum of two instances within a shard.
	
	\item {\emph{Shard Limit 1}}: This is the most stringent target, demanding that each node in a shard has a unique value for every characteristic,  enforcing the highest level of decentralization.
\end{itemize}
As our analysis in the appendix shows, there is a limited decentralization in terms of the country characteristic. For that reason, we analyze an additional target  that combines elements of the Shard Limit 2 and Shard Limit 1 targets. This target, named \emph{Hybrid Shard Limit}, enforces a shard limit of 3 for the country characteristic in larger subnets (with 28 nodes or more), a shard limit of 2 for the country characteristic in all other subnets, and a shard limit of 1 for all other characteristics. 

\begin{figure}[ht]
	\centering
	\includegraphics[width=0.8\textwidth]{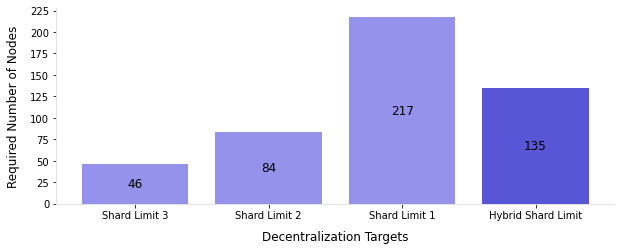}
	\caption{Required number of nodes for different decentralization targets.}
	\label{bar_chart_results}
\end{figure}

Figure \ref{bar_chart_results} shows the number of additional nodes required for the analyzed decentralization targets. As anticipated, the strictness of the decentralization target is directly proportional to the increase in the number of required nodes. The target {Hybrid Shard Limit}  was ultimately selected and approved by the Internet Computer Protocol (ICP) community \cite{ICProposalNov2023}, as it constitutes a healthy balance between decentralization and resource usage.
\vspace{10pt}
\vspace{10pt}

\begin{figure}
	\centering
	\captionsetup{justification=centering}
	\includegraphics[width=1\textwidth]{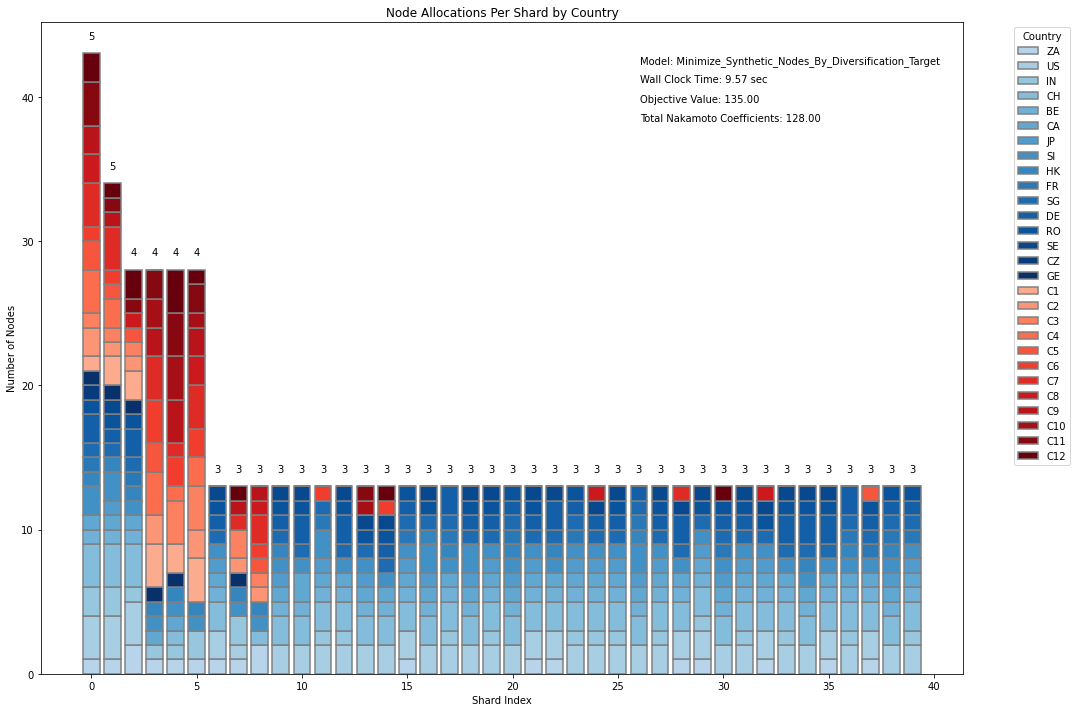}
	\caption{Node allocation to shards under the hybrid shard limit.}
	\label{decentralization_target_hybrid_country}
\end{figure}

Figure~\ref{decentralization_target_hybrid_country} illustrates the allocation of nodes to shards in relation to the country characteristic\footnote{While the optimization considers all characteristics simultaneously, due to the two-dimensional nature of our figures, each figure can only visualize the allocation per one characteristic.},  for the {Hybrid Shard Limit} target. 
Each bar in the figure represents a shard, arranged in descending order of size from left to right. The color segments within each bar indicate nodes from the same country, with blue shades denoting existing nodes and red shades representing potential additions from 12 new countries. Atop each bar, the displayed number indicates the  Nakamoto coefficient.% for the shard, pertaining to the country characteristic.

%\vspace{10pt}
%\vspace{10pt}

\clearpage

\section{Conclusion and Future Directions}
We introduced an optimization framework that provides a quantitative method for modeling diverse decentralization characteristics, to express the target topology for a sharded blockchain network, and to identify the minimal set of nodes that satisfies that topology.  This approach facilitates algorithmic decision-making for node onboarding and offboarding, striking a balance between achieving diversification and maintaining economic viability within blockchain networks. 
Our approach depends on the availability of reliable characteristics of nodes, such as node ownership, geographical location, or jurisdiction. Recognizing this as a limitation of our proposed methodology, we discussed methods to collect and enhance the trustworthiness of node characteristics. More sophisticated solutions to further increase their reliability is subject to future work.

The paper demonstrates the practical use of this framework by reporting on how the ICP community has analyzed and agreed on a target topology. To the best of our knowledge, this is the first time that such an optimization framework has been presented and applied to a blockchain network.
It is worth mentioning that a blockchain network without shards is essentially a network with a single shard. Therefore, the results presented here can also be applied to blockchains without shards. 

We see various directions of future work. For example, given an existing allocation of nodes to shards and a target topology, what are the minimal changes to the current allocation of nodes that facilitate reaching the target topology. In addition, how does the decision to include a subset of the total nodes required for the target topology impact the set of remaining nodes that will still be required? 
Another direction of future work may focus on the security implications resulting from the quantitative nature of the model. An adversary may try to control a shard by deducing from the model which kinds of nodes must be added in order for the nodes to be allocated to the targeted shard. How can such attacks be prevented? 

We also have various ideas for refining the model to incorporate additional aspects into the analysis. Besides taking further characteristics into consideration, one could weigh characteristics against one another more explicitly. For example, is it better to be more decentralized in terms of data center ownership or the country assignments? The cost function may also be refined. For example, it may take the ecological impact of nodes into consideration.

We believe that the quantitative approach presented in this paper allows for a more objective discussion of how much decentralization is needed for specific blockchain applications. This allows to reconcile the growing need for secure systems with a limited pool of resources.

%\appendix
%define the following sections to hide their Section Number (Notes Style)
\ledgernotes

%define the following sections to have the Appendix Style
%\appendix

\section*{Acknowledgements}
The authors would like to extend their sincere gratitude to Björn Tackmann, Sven Fischer, and Saša Tomić for 
 invaluable discussions  during the development of this model. 
\newpage

\section{Appendix A: Initial assessment of node decentralization}
\label{app:initial_assessment}
Given the provided model input in Section~\ref{sec:utilized_model_inputs}, we conduct a preliminary evaluation of node decentralization on the Internet Computer, focusing on four distinct characteristics:
\begin{itemize}
	\item {\emph{Node owner}} describes which entity, business or individual, owns and ultimately controls the node,
	\item {\emph{Data center}} specifies the location and building/campus where the node is running determining the network connectivity and power provider used, 
	\item {\emph{Data center provider}} defines the business entity controlling the data center, and
	\item {\emph{Country}} is the country where the data center is located, implying the applicable jurisdiction. 
\end{itemize}
For optimal decentralization, each shard would only have unique representations of each characteristic. For example, within a shard, no two nodes should be owned by the same node owner. To visualize this  we use a \emph{node topology matrix} that looks at one characteristics:
\begin{itemize}
	\item Columns represent shards,
	\item Rows represent values of the considered characteristic,
	\item Cells with a cross denote that the corresponding shard, i.e. column, requires a node with this characteristic value,
	\item Cells that are colored denote that a node with this characteristic value is available for the corresponding shard, i.e. column.
\end{itemize}
Every node can only be mapped to one shard. We require that rows represent characteristic values in ascending order from top to bottom, while shards, i.e. columns, are listed sorted by their size in descending order from left to right.

\begin{figure}
	\captionsetup{belowskip=10pt}
	\centering
	\captionsetup{justification=centering}
	\includegraphics[width=1\textwidth]{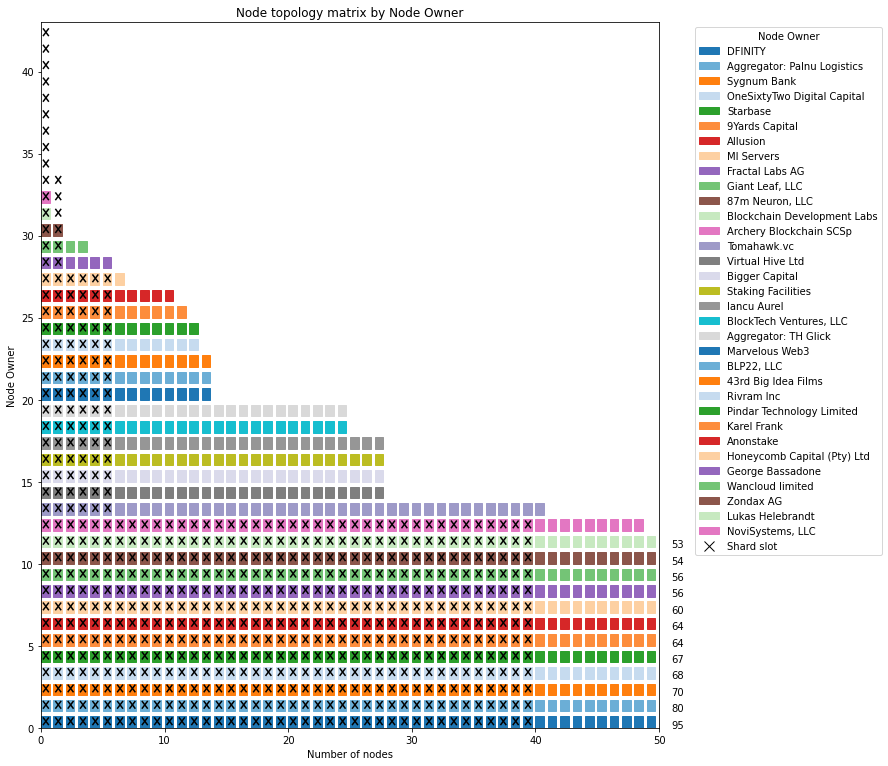}
	\caption{Node topology matrix for the “node owner” characteristic.}
	\label{topology_matrix_node_provider}
\end{figure}

Figure~\ref{topology_matrix_node_provider} shows the node topology matrix for the characteristic \emph{node owner}. One can infer that achieving near-optimal decentralization is plausible. Notably, only 13 shard slots, visible as cells with crosses that are not colored, remain vacant in the two largest shards on the left.

Additionally, it becomes clear that several node owners have a disproportionately high number of nodes compared to the total number of shards required. For achieving optimal decentralization across node owners, owners should not operate more nodes than the number of required shards. Otherwise, either there are shards where node owners operate more than one node or there are nodes that remain unassigned, i.e. not used. 
\clearpage

\begin{figure}
	\centering
	\captionsetup{justification=centering}
	\includegraphics[width=0.7\textwidth]{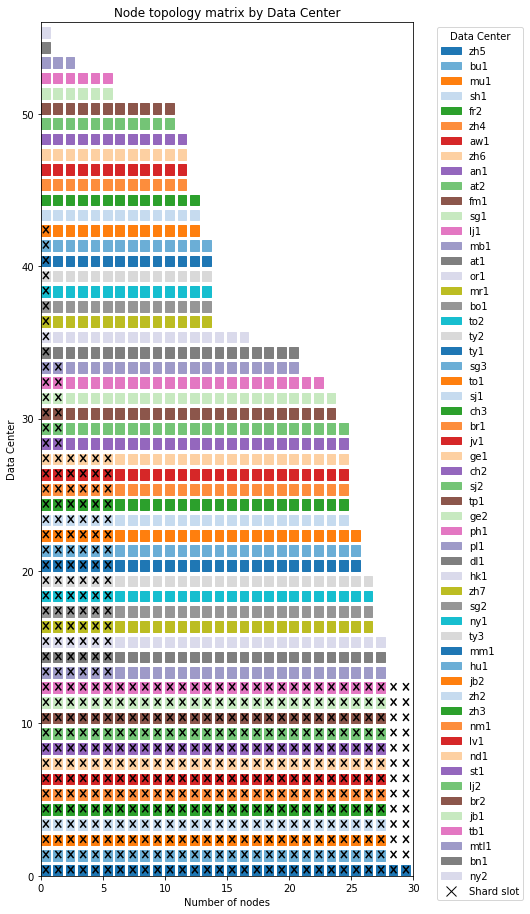}
	\caption{Node topology matrix for the “data center” characteristic.}
	\label{topology_matrix_data_center}
\end{figure}

Figure~\ref{topology_matrix_data_center}  shows the node topology matrix for the  characteristic \emph{data center}. It is evident that the Internet Computer can be fully  decentralized with respect to the characteristic data center. Notably, even the largest shard, comprising 43 nodes, can be covered accommodating each data center only once. 

It is worth pointing out that the crosses in the bottom right of Figure~\ref{topology_matrix_data_center}, which do not overlay colored cells, do not present a decentralization concern. These available slots can readily be populated using unassigned nodes (cells without crosses) from the upper rows without compromising the decentralization objective. 

\begin{figure}
	\centering
	\captionsetup{belowskip=10pt, justification=centering}

	\includegraphics[width=0.8\textwidth]{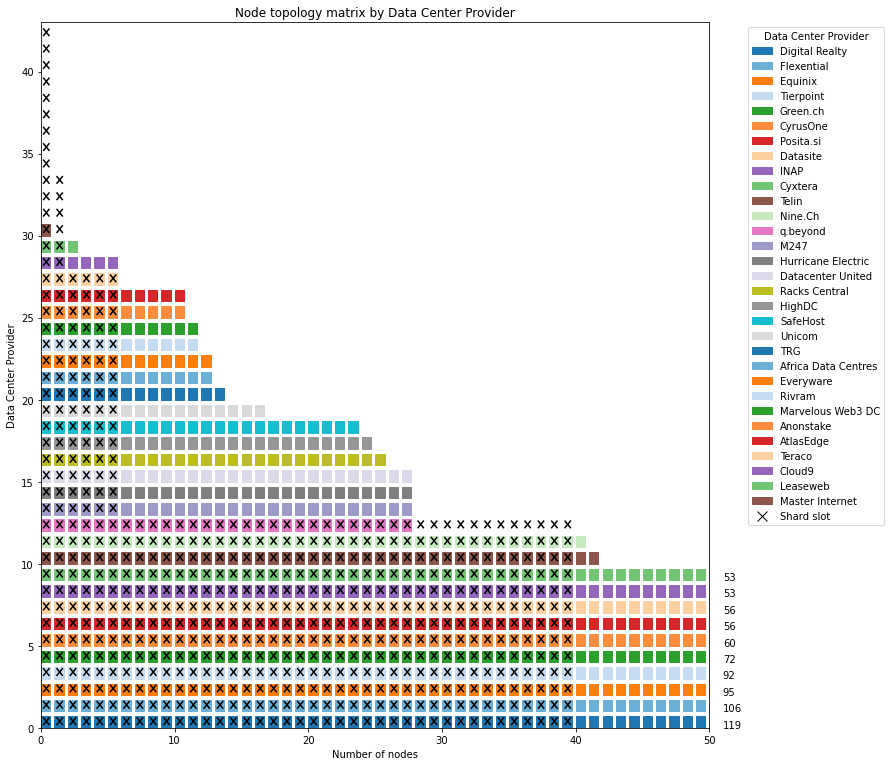}
	\caption{Node topology matrix for the “data center provider” characteristic.}
	\label{topology_matrix_data_center_provider}
\end{figure}

Figure~\ref{topology_matrix_data_center_provider}  shows the node topology matrix for the  characteristic \emph{data center provider}. It shows a similar level of decentralization as the characteristic node owner. 

\begin{figure}
	\centering
	\captionsetup{belowskip=10pt, justification=centering}
	
	\includegraphics[width=1\textwidth]{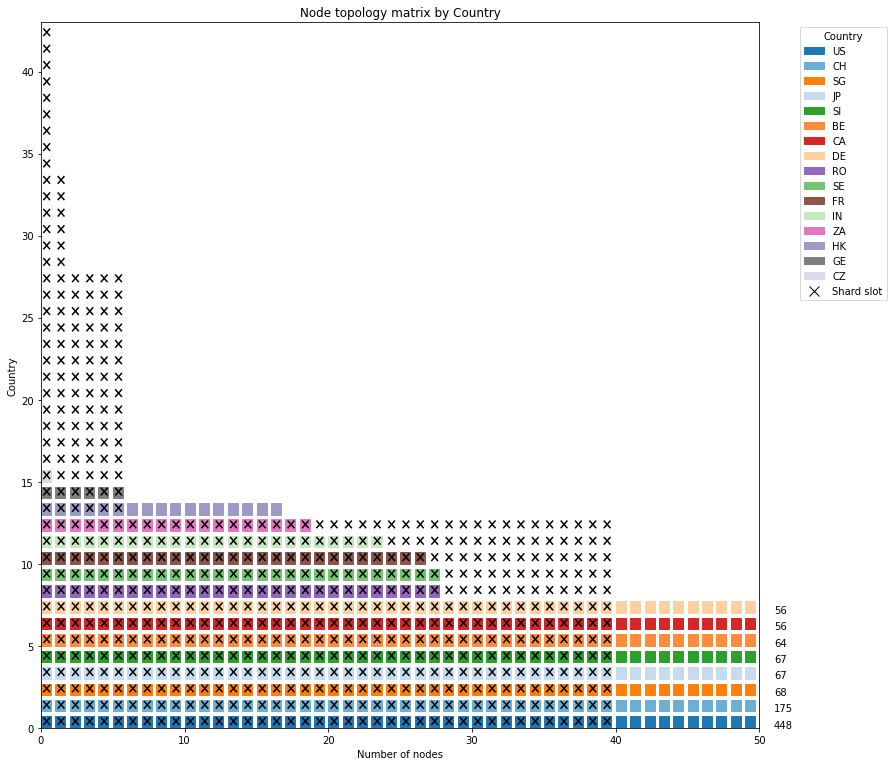}
	\caption{Node topology matrix for the “country” characteristic.}
	\label{topology_matrix_country}
\end{figure}

Figure~\ref{topology_matrix_country} shows the node topology matrix for the  characteristic \emph{country}. A stark contrast is evident when compared to the previous matrices.
Currently, nodes are operated in 15 different countries. Given this situation, it is infeasible to cover the large shards on the left without either adding nodes in significantly more countries or weakening the constraint that each country is represented only once within a shard. This challenge is highlighted by the 50+ cells with crosses and no coloring.

\newpage
Even to cover  some of the shards with 13 nodes, there are not sufficiently many diverse nodes. This shortcoming is visualized by the 50+ cells in the middle of the matrix that contain a cross but are not colored. This example illustrates how node topology matrices are a useful tool to diagnose decentralization constraints and where they are satisfied or remain unsatisfied.

\begin{figure}
	\centering
	\captionsetup{justification=centering}
	\includegraphics[width=1\textwidth]{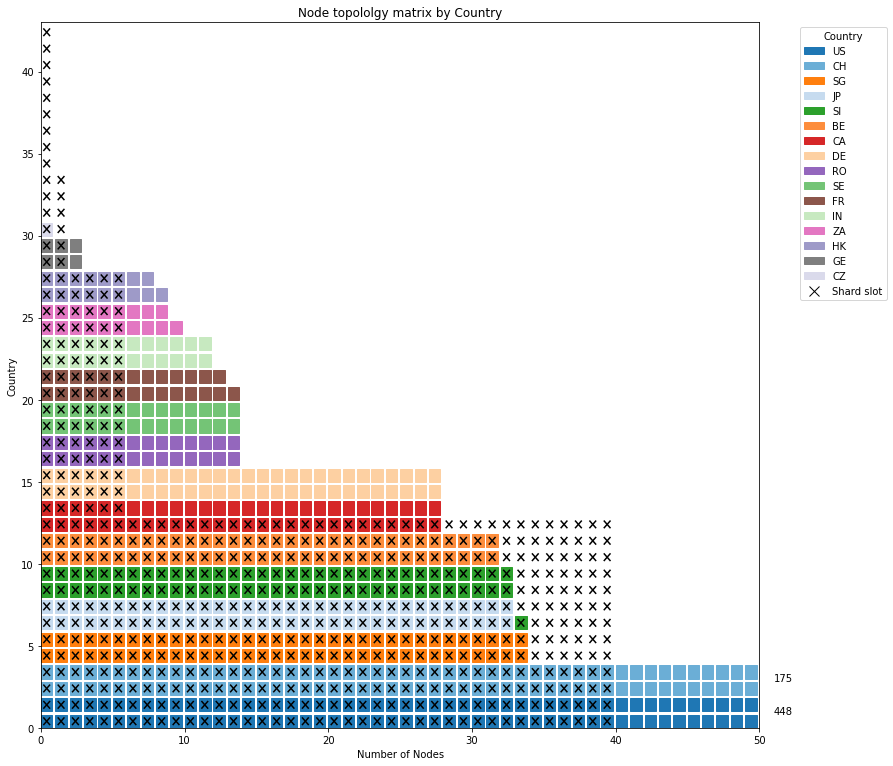}
	\caption{Node topology matrix for the “country” characteristic with two rows per country.}
	\label{topology_matrix_country2}
\end{figure}

To address the limited country diversity, one could lower the decentralization requirement per country. For example, one could  specify that every country should, at most, be represented twice (instead of once) within a shard. Implementing this modification results in the  node topology matrix shown in Figure~\ref{topology_matrix_country2}, where we permit two rows for each country.

\clearpage

\section{Appendix B: Visualization of Model Results}
In this section, we present additional visualizations corresponding to the decentralization targets described in Section \ref{sec:model_results}. Please also refer to Section  \ref{sec:model_results} for a description of the structure of the visualizations.

\begin{figure}
	\captionsetup{belowskip=10pt}
	\centering
	\captionsetup{justification=centering}
	\includegraphics[width=1\textwidth]{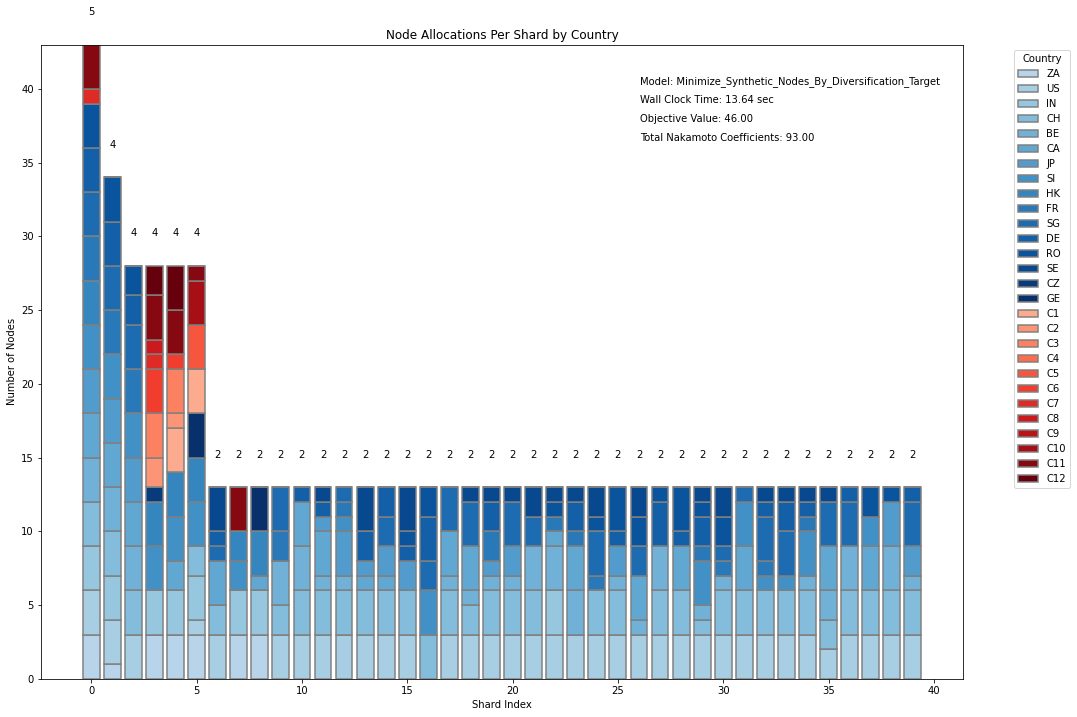}
	\caption{Node allocation to shards under Shard Limit 3.}
	\label{decentralization_target_shard_limit_3}
\end{figure}

\emph{Results Shard Limit 3}: Under this decentralization target, the model determines that 46 additional nodes are required.  Figure~\ref{decentralization_target_shard_limit_3} shows the node allocation to shards with respect to the characteristic country. 

\begin{figure}
	\centering
	\captionsetup{justification=centering}
	\includegraphics[width=1\textwidth]{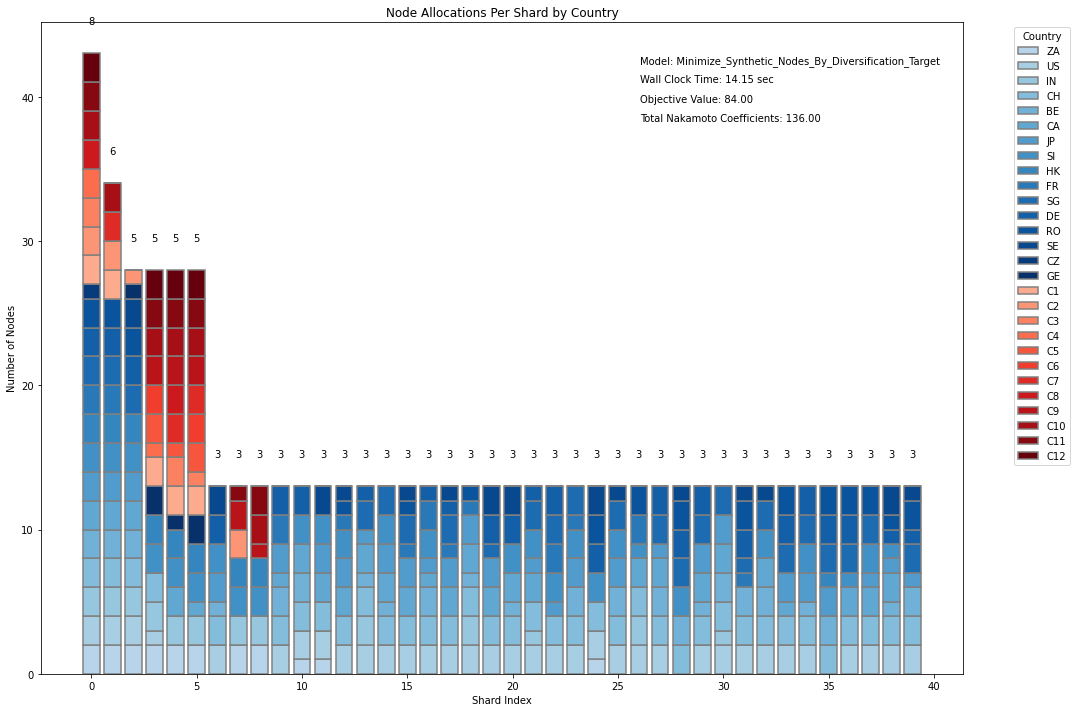}
	\caption{Node allocation to shards under Shard Limit 2.}
	\label{decentralization_target_shard_limit_2}
\end{figure}

\emph{Results Shard Limit 2}: For the stricter decentralization target, enforcing a shard limit of 2 for all characteristics, an additional 84 nodes are required. Figure~\ref{decentralization_target_shard_limit_2} shows that additional nodes are mainly required for the larger shards on the left (shard 1, 2), as well as for the generation 2 shards (shard number 4, 5, 6, 8, 9).

\begin{figure}
	\centering
	\captionsetup{justification=centering}
	\includegraphics[width=1\textwidth]{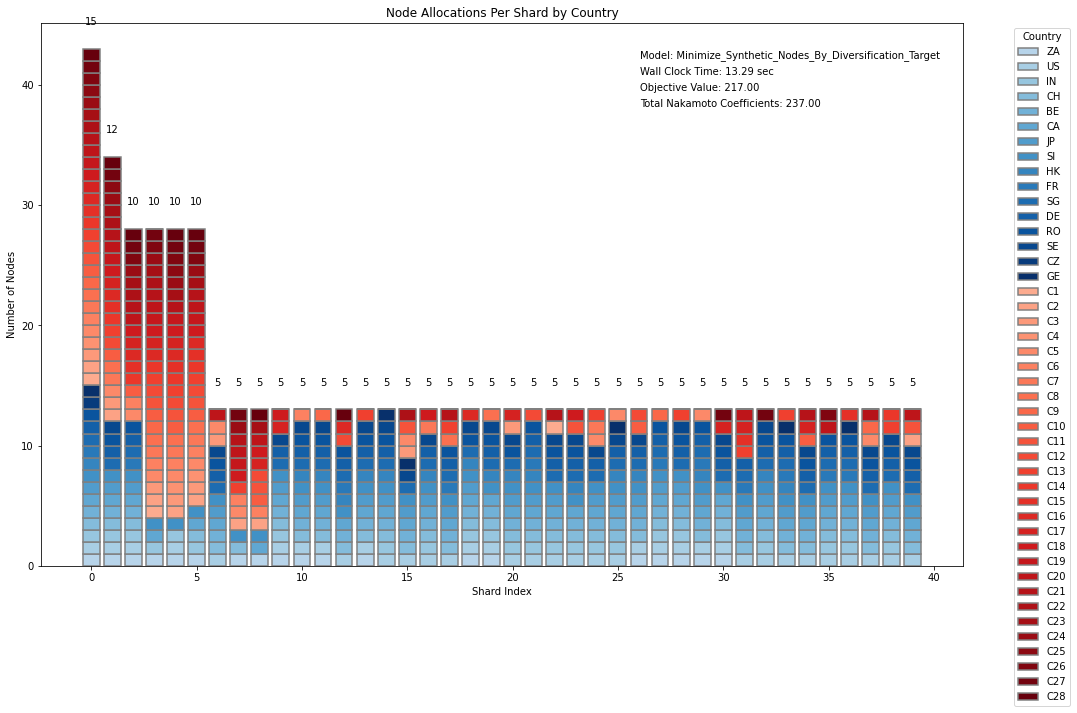}
	\caption{Node allocation to shards under Shard Limit 1.}
	\label{decentralization_target_shard_limit_1}
\end{figure}

\emph{Results Shard Limit 1}: For achieving maximal decentralization across all characteristics 217 additional nodes are required, sourced from 28 new countries, as depicted in Figure~\ref{decentralization_target_shard_limit_1}.

\begin{figure}
	\centering
	\captionsetup{justification=centering}
	\includegraphics[width=1\textwidth]{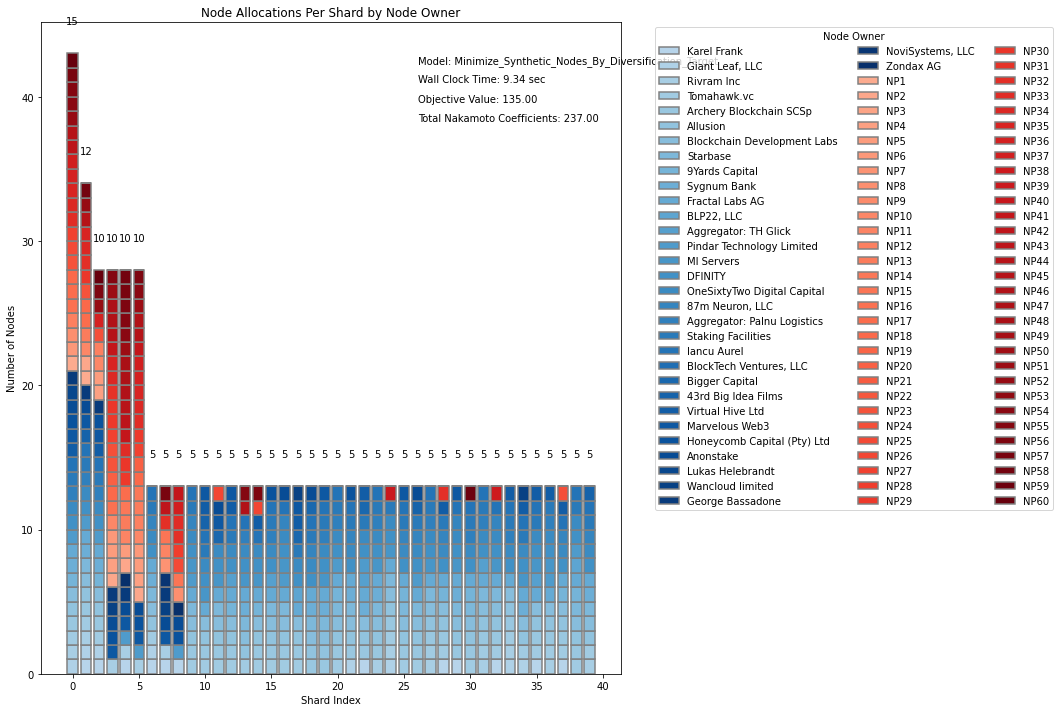}
	\caption{Node allocation to shards under the Hybrid Shard Limit.}
	\label{decentralization_target_hybrid}
\end{figure}

\emph{Results Hybrid Shard Limit}: This target implies maximal decentralization in terms of node owners, data centers and data center providers, while allowing a sub-optimal decentralization in terms of the characteristic country. As shown in Figure~\ref{decentralization_target_hybrid}, every shard has the ideal Nakamoto Coefficient, e.g. coefficient 5 for all 13-node shards, with respect to the characteristic node owner. For this level of decentralization 135 additional  nodes are required.  	

\clearpage  % Ensure all figures are processed before starting the bibliography

%%%%%%%%%%%%%%%%%%%%%%%%%%%%%%%%%%%%%%%%%%%%

%define the following sections to hide their Section Number (Notes Style)
\ledgernotes

%\section{Author Contributions}

%State the contribution made by each author.  Refer to authors using their initials, for example, ``FAA developed the code to perform the simulation (65\%) and SBA analyzed the results (35\%).  They both contributed equally to manuscript preparation.''

%AUTHOR: comment out if using thebibliography
%\theendnotes

%\newpage
\bibliographystyle{ledgerbib}
\bibliography{tempbib}

%AUTHOR: comment out, this is used to make sure the Creative Commons License
%image fits on page
%\newpage

%AUTHOR: this is an alternative to using Endnotes
%\bibliographystyle{nature}

%\begin{thebibliography}{99}

%\bibitem{itemName}
%Last, F.
%\newblock {``Title.''}
%\newblock {\em Publication} (XX Month 201X)
%  \url{http://}

%\end{thebibliography}

%add the Creative Commons license footer to the last page
\thispagestyle{pagelast}

\end{document}